\documentclass[%
preprint,
 amsmath,amssymb,
 aps,
]
{revtex4-1}
\usepackage{graphicx}
\usepackage{braket}
\usepackage{bm}
\usepackage{amsmath}
\usepackage{color}
\usepackage{ulem}
\begin{document}
\title{Erosion of $\bm{N=28}$ shell closure: Shape coexistence and monopole transition}
\author{Y. Suzuki}
\affiliation{Department of Physics, Hokkaido University, 060-0810 Sapporo, Japan}
\author{W. Horiuchi}
\email{whoriuchi@nucl.sci.hokudai.ac.jp}
\affiliation{Department of Physics, Hokkaido University, Sapporo 060-0810, Japan} 
\author{M. Kimura}
\email{masaaki@nucl.sci.hokudai.ac.jp}
\affiliation{Department of Physics, Hokkaido University, Sapporo 060-0810, Japan} 
\affiliation{Nuclear Reaction Data Centre, Hokkaido University, Sapporo 060-0810, Japan} 
\affiliation{RIKEN Nishina Center, Wako, Saitama 351-0198, Japan} 
\date{\today}

\begin{abstract}
  \noindent{\bf Background}: In neutron-rich nuclei neighboring $^{42}{\rm Si}$, the quenching of
 the  $N=28$ shell gap occurs and is expected to induce the shape coexistence  in their excitation 
 spectra.\\ 
 \noindent{\bf Purpose}: We show that different nuclear shapes coexist in $N=28$ isotones $^{40}{\rm
 Mg}$, $^{42}{\rm Si}$, and $^{44}{\rm S}$, and investigate observables to probe it.\\  
 \noindent {\bf Method}: Antisymmetrized molecular dynamics with Gogny D1S density functional is
 applied to describe the shape coexistence phenomena  without {\it ad hoc} assumption of the nuclear
 shape.\\ 
 \noindent {\bf Results}: We find that rigid shapes with different deformations coexist in  the
 ground and the first excited $0^+$ states of $^{40}{\rm Mg}$ and $^{42}{\rm Si}$, while in
 $^{44}{\rm S}$ the states exhibit large-amplitude collective motion, which does not have any
 particular shape. These characteristics are reflected well in the  monopole transition strengths.\\ 
 \noindent{\bf Conclusion}: The quenching of the $N=28$ shell gap leads to the unique shape
 coexistence in the $N=28$ isotones, which can be probed by the monopole transition strengths.
\end{abstract}

\maketitle

\section{introduction}

The Fermi surface of neutron-rich unstable nuclei exhibits a different characteristics from these of
stable nuclei due to the quenching of the magic shell
gap~\cite{Wilkinson1959,Talmi1960,Thibault1975,Sorlin2008}. 
Various exotic phenomena caused by the quenching  have been observed, such as halo structure
~\cite{Tanihata1985,Ozawa2001,Horiuchi2006,Nakamura2009,Takechi2012,Horiuchi2012,Minomo2012} and
evolution of nuclear
shape~\cite{Warburton1990,Fukunishi1992,Motobayashi1995,Sorlin1993,Scheit1996,Glasmacher1997}, which
have been  major topics of interest in modern nuclear physics. 

The neutron number $N=28$ is the smallest magic number  whose shell gap, i.e., the energy gap between
the $f_{7/2}$ and $p_{3/2}$ orbits, is produced by the spin-orbit splitting.  The quenching of this
shell gap in the neutron-rich $N=28$ isotones ($^{40}{\rm Mg}$, $^{42}{\rm Si}$ and $^{44}{\rm S}$)
induces the neutron quadrupole collectivity because these neutron orbits have the same parity and the
angular momenta different by two units. Furthermore, these isotones have a similar quadrupole
symmetry also in the proton Fermi levels, i.e., the half-filling of the $sd$-shells.
Therefore, the strong quadrupole collectivity of both protons and neutrons in these isotopes is
expected, leading to large quadrupole deformations of the ground states. Experimental evidences
include the reduction of the $2^+_1$ state
energy~\cite{Scheit1996,Glasmacher1997,Hartmann2002,Sohler2002,Nowak2016,Crawford2019} and the 
ratio of the excitation energies of the $2^+_1$ and $4^+_1$ states~\cite{Takeuchi2012}; and the
enhancement of the electric-quadrupole transition strength~\cite{Hartmann2002,Scheit1996}. A number
of nuclear model calculations have also suggested various
deformations in this mass region~\cite{Delaroche2010,Rodriguez2011,Kimura2013,Egido2016,Tsunoda2020}.  

Another consequence of the quenching of the $N=28$ shell gap might be the shape coexistence; the
existence of the low-lying $0^+_2$ or non-yrast states which have a shape different from that of the 
ground state~\cite{Heyde2011}. Experimentally, many non-yrast states have already been observed in
$^{44}{\rm S}$ at small excitation energies below 4 MeV, implying the shape
coexistence~\cite{Santiago-Gonzalez2011,Utsuno2015}.  Several theoretical studies have already
discussed the shape coexistence~\cite{Santiago-Gonzalez2011,Utsuno2015,Kimura2016}, but how
the nuclear shape affects the $N=28$ shell closure is not understood clearly. Furthermore,
observables that can appropriately reflect different nuclear shapes are not known.

Here, we study the structure of $N=28$ isotones by using the antisymmetrized molecular dynamics
(AMD)~\cite{Kanada-Enyo2003,Kanada-Enyo2012,Kimura2016}, which can describe various nuclear
deformations without {\it ad hoc} assumption.
In Ref.~\cite{Suzuki2021}, two of the present authors (Y.S. and M.K.)
investigated the ground-state deformation of $N=28$ isotones and reported the isotope dependence of
the deformation and importance of the triaxial deformation to explain the observed data.
In this work, we study the excitation spectra of $N=28$ isotones with a special emphasis on the
shape coexistence in the ground and 
$0_2^+$ states and its relationship with monopole transitions. 

We found the coexistence
of prolately- and oblately-deformed states in $^{40}{\rm Mg}$, and the coexistence of
oblately-deformed and spherical states in $^{42}{\rm Si}$. Differently from these nuclei, 
$^{44}{\rm S}$ has large shape fluctuation and does not have definite shape, which can be 
regarded as large amplitude collective motion.  We also found that the structure of the neutron
Fermi surface, i.e., the  degree of $N=28$ shell gap and the  ordering of the neutron orbits, is
strongly dependent on the nuclear shape. In the prolately-deformed states, the $N = 28$ 
shell gap disappears due to the inversion of the neutron orbits near the Fermi level. On the other
hand, in the oblately-deformed states, the shell gap is kept large, but the mixing of $f$- and
$p$-wave in the neutron orbits close to the Fermi surface erodes the $N=28$ shell closure. We show
that the monopole transition is a promising observable that can measure the shape coexistence and
the neutron orbits near the Fermi level. 

This paper is organized as follows. In the next section, we briefly explain the framework of AMD. In 
Sec.~\ref{sec:result}, we present the numerical results and investigate the shape coexistence
phenomena in the $N=28$ isotones. We show that the monopole transition between the ground and
$0^+_2$ states can be a probe for the shape coexistence. Sec.~\ref{sec:summary} summarizes this
work.

\section{Theoretical framework}

Here, we briefly describe the AMD framework. As the setup is the same, see Ref.~\cite{Suzuki2021}
for more details. We use a microscopic Hamiltonian given by 
\begin{align}
 H=\sum_{i}^{A} t_{i}-t_\mathrm{cm}
 +\frac{1}{2}\sum_{ij}^{A} v_{ij}^\mathrm{NN}
 +\frac{1}{2}\sum_{ij}^{A} v_{ij}^{\mathrm{C}},
\end{align}
where $t_i$ is the kinetic energy term of the $i$th nucleon with the center-of-mass energy term 
$t_{\rm cm}$ being subtracted.  The Gogny D1S density functional~\cite{Berger1991} is employed 
for the nucleon-nucleon interaction $v_{ij}^{\rm NN}$,
and $v_{ij}^{\mathrm{C}}$ denotes the Coulomb interaction.
The variational wave function is a parity-projected Slater 
determinant
\begin{equation}
\Phi^{\pi}=P^\pi \mathcal{A} \{\varphi_{1}\varphi_{2}\dots\varphi_{A}\},
\end{equation}
where $P^\pi$ is the parity projection operator.
The single-particle wave packet $\varphi_{i}$
 is taken as a deformed Gaussian form~\cite{Kimura2004a}
\begin{align}
 \varphi_i(\bm{r}) &= \exp
 \set{-\sum_{\sigma=x,y,z} {\nu_\sigma} \left (r_{\sigma}-Z_{i\sigma} \right)^2}
 \chi_{i} \eta_{i}
\end{align}
with the spin and isospin functions
\begin{align}
\chi_{i}&=a_{i}\chi_\uparrow+b_{i}\chi_\downarrow, \quad
 \eta_{i}= \set{\mathrm{proton\ or\ neutron}}.
\end{align}
The variational parameters are the Gaussian width $\nu_{x}, \nu_{y}, \nu_{z}$ and their centroids
$\bm Z_i$; and spin direction $a_i$ and $b_i$. 
They are determined by the energy variation with the
constraint on the matter quadrupole deformation parameters
$\beta$ and $\gamma$~\cite{Kimura2012},
which yields the optimized wave function $\Phi^{\pi}(\beta , \gamma)$ for given values of $\beta$
and $\gamma$.
The value of $\beta$ and $\gamma$ is chosen on the triangular grid in the
$\beta$-$\gamma$ plane ranging $0\leq\beta\leq 0.6$ and $0\leq\gamma\leq60^\circ$. The side of the
triangular grid is chosen to be 0.05.  

These optimized wave functions are projected
to the eigenstate of the angular momentum, and they are 
superposed (generator coordinate method; GCM~\cite{Hill1953})
to describe the nuclear shape fluctuation
\begin{align}
 \Psi_{M\alpha}^{J\pi}
 =\sum_{iK}  g_{iK\alpha} P^J_{MK}\Phi^\pi (\beta_i , \gamma_i),\label{eq:gcmwf}
\end{align}
where $\beta$ and $\gamma$ are the generator coordinates, and $P^J_{MK}$ denotes the angular
momentum projector. The coefficients $g_{iK\alpha}$ and eigenenergies $E_\alpha$ are obtained by
solving the Hill-Wheeler equation~\cite{Hill1953}
\begin{align}
 & \sum_{j K'} H_{iKj K'} g_{jK'\alpha}
  =E_\alpha\sum_{jK'}N_{iKjK'}g_{j K'\alpha}\label{eq:HWg}
\end{align}
with
\begin{align}
 &H_{iKjK'}=\braket {P^J_{MK}\Phi^\pi(\beta_i,\gamma_i)|
 H|P^J_{MK'}\Phi^\pi(\beta_j,\gamma_j)},\\
 &N_{iKjK'}=\braket {P^J_{MK}\Phi^\pi(\beta_i,\gamma_i)|
 P^J_{MK'}\Phi^\pi(\beta_j,\gamma_j)}.
\end{align}

To analyze
the properties of the ground and $0^+_2$ states, we calculate the following
quantities. The first is the energy surface and GCM amplitude. The energy surface is the energy of
the wave function projected to the $J^\pi=0^+$ state with quadrupole deformation $\beta$ and $\gamma$ 
\begin{align}
 E(\beta,\gamma) = \frac{\braket{P^J_{MK}\Phi^\pi(\beta,\gamma)|H|P^J_{MK}\Phi^\pi(\beta,\gamma)}}
 {\braket{P^J_{MK}\Phi^\pi(\beta,\gamma)|P^J_{MK}\Phi^\pi(\beta,\gamma)}}.
\end{align}
It describes how the binding energy changes as a function of $\beta$ and $\gamma$.
The GCM amplitude is defined as the overlap between the GCM wave function 
$\Psi^J_{M\alpha}$ [Eq.~(\ref{eq:gcmwf})]
and the basis wave functions $P^J_{MK}\Phi^\pi(\beta,\gamma)$  
\begin{align}
 f(\beta,\gamma) = \frac{\braket{\Psi^{J\pi}_{M\alpha}|P^J_{MK}\Phi^\pi(\beta,\gamma)}}
 {\sqrt{\braket{P^J_{MK}\Phi^\pi(\beta,\gamma)|P^J_{MK}\Phi^\pi(\beta,\gamma)}}}.
\end{align}
The value of $\beta$ and $\gamma$ that give the maximum amplitude of $|f(\beta,\gamma)|$ 
is roughly regarded as the equilibrium shape of the GCM wave function.
In this sense, the distribution of
$|f(\beta,\gamma)|$ indicates the shape fluctuation around the equilibrium shape.  

The second is the single-particle energies and orbits. To evaluate them, we introduce the
orthonormalized single-particle wave functions by a linear transformation of the single-particle
wave packets
\begin{align}
\widetilde{\varphi}_p
=\frac{1}{\sqrt{\mu_p}} \sum_{i} c_{ip} \varphi_{i},
\end{align}
where $\mu_p$ and $c_{ip}$ are the eigenvalues and the eigenvectors of the overlap matrix 
$B_{ij} = \braket{\varphi_i|\varphi_j}$. The single-particle Hamiltonian is defined as~\cite{Dote1997}
\begin{align}
h_{pq} &= \braket{\widetilde{\varphi}_{p} | t | \widetilde{\varphi}_{q}}
+\sum_{r} \braket{\widetilde{\varphi}_{p} \widetilde{\varphi}_{r} |
 v^{\mathrm{NN}} + v^{\mathrm{C}} |
 \widetilde{\varphi}_{q} \widetilde{\varphi}_{r} - \widetilde{\varphi}_{r} \widetilde{\varphi}_{q}
 } \nonumber\\ 
 &+ \frac{1}{2} \sum_{r, s} \braket{ \widetilde{\varphi}_{r} \widetilde{\varphi}_{s} 
| \widetilde{\varphi}_{p}^{\ast} \widetilde{\varphi}_{q} \frac{\delta v^{\mathrm{NN}}}{\delta \rho} |
\widetilde{\varphi}_{r} \widetilde{\varphi}_{s} - \widetilde{\varphi}_{s} \widetilde{\varphi}_{r}}.
\end{align} 
The eigenvalues and eigenvectors of $h_{pq}$ give the single-particle energies and wave functions
in the present AMD approach.

\section{results and discussion}\label{sec:result}
  \subsection{Shape coexistence and erosion of the $\bm{N=28}$ shell closure}

\begin{figure*}[ht]
\centering
\includegraphics[width=0.8\hsize]{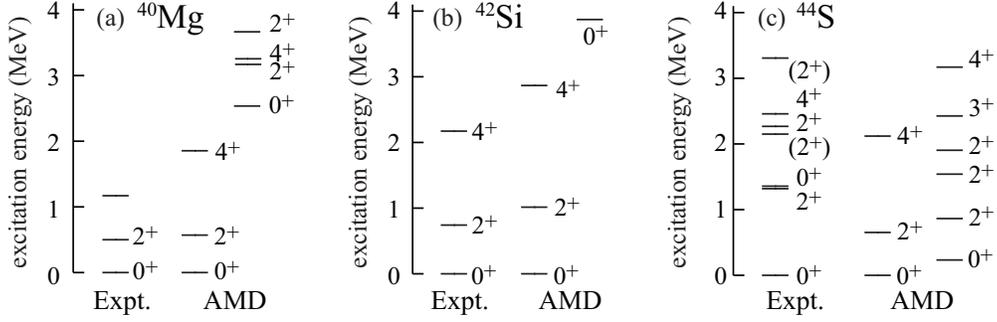}
\caption{ Excitation energies of yrast and non-yrast states of (a) $^{40}{\rm Mg}$,
 (b) $^{42}{\rm Si}$ and (c) $^{44}{\rm S}$. Experimental data is taken from
 Refs.~\cite{Sohler2002,Crawford2019,Takeuchi2012}.}
 \label{fig:level} 
\end{figure*}

\begin{figure}[ht]
\centering
\includegraphics[width=0.5\hsize]{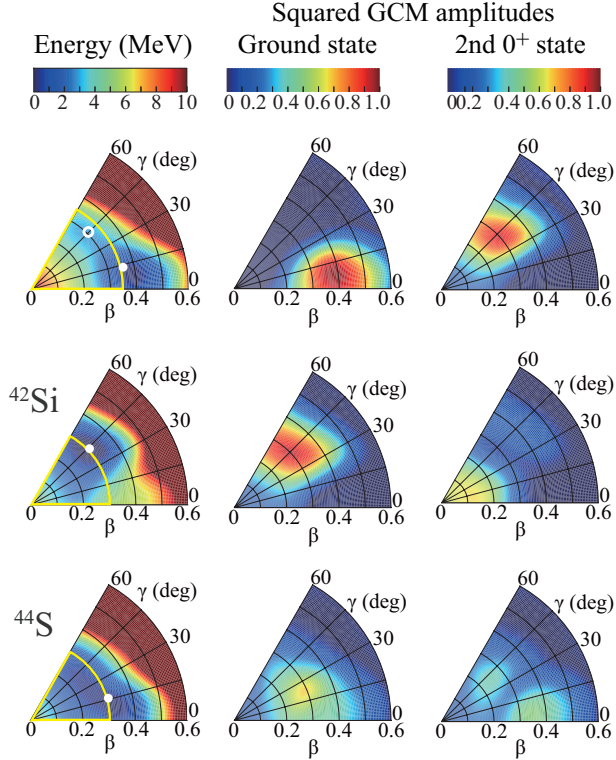}
\caption{Energy surfaces and GCM amplitudes of the ground and $0^+_2$ states of the $N=28$ isotones 
  as a function of the quadrupole deformation parameters  $\beta$ and $\gamma$. The panels in the
 left column draws the energy surfaces of the $J^{\pi}=0^{+}$ states, in which filled (open) circles
 indicates the position of the (local) energy minima. Yellow-colored sector line shows the path along
 which we plot the GCM amplitude and neutron single-particle energies in Figs.~\ref{fig:amp} and
 \ref{fig:spl}. The panels in the middle and right columns show the squared GCM amplitudes for
 the 
 ground and $0^+_2$  states, respectively.} 
 \label{fig:surface} 
\end{figure}

Figure~\ref{fig:level} shows the excitation spectra of $^{40}{\rm Mg}$, $^{42}{\rm Si}$ and
$^{44}{\rm S}$ obtained by the present calculation.  We find that the ground and $0^+_2$ states of
these nuclei coexist at small excitation energies less than 4 MeV, which exhibits shape
coexistence. Here, we explain the shape of each nucleus based on its energy surface and squared GCM
amplitude shown in  Fig.~\ref{fig:surface}.


\begin{figure}[ht]
\centering
\includegraphics[width=0.5\hsize]{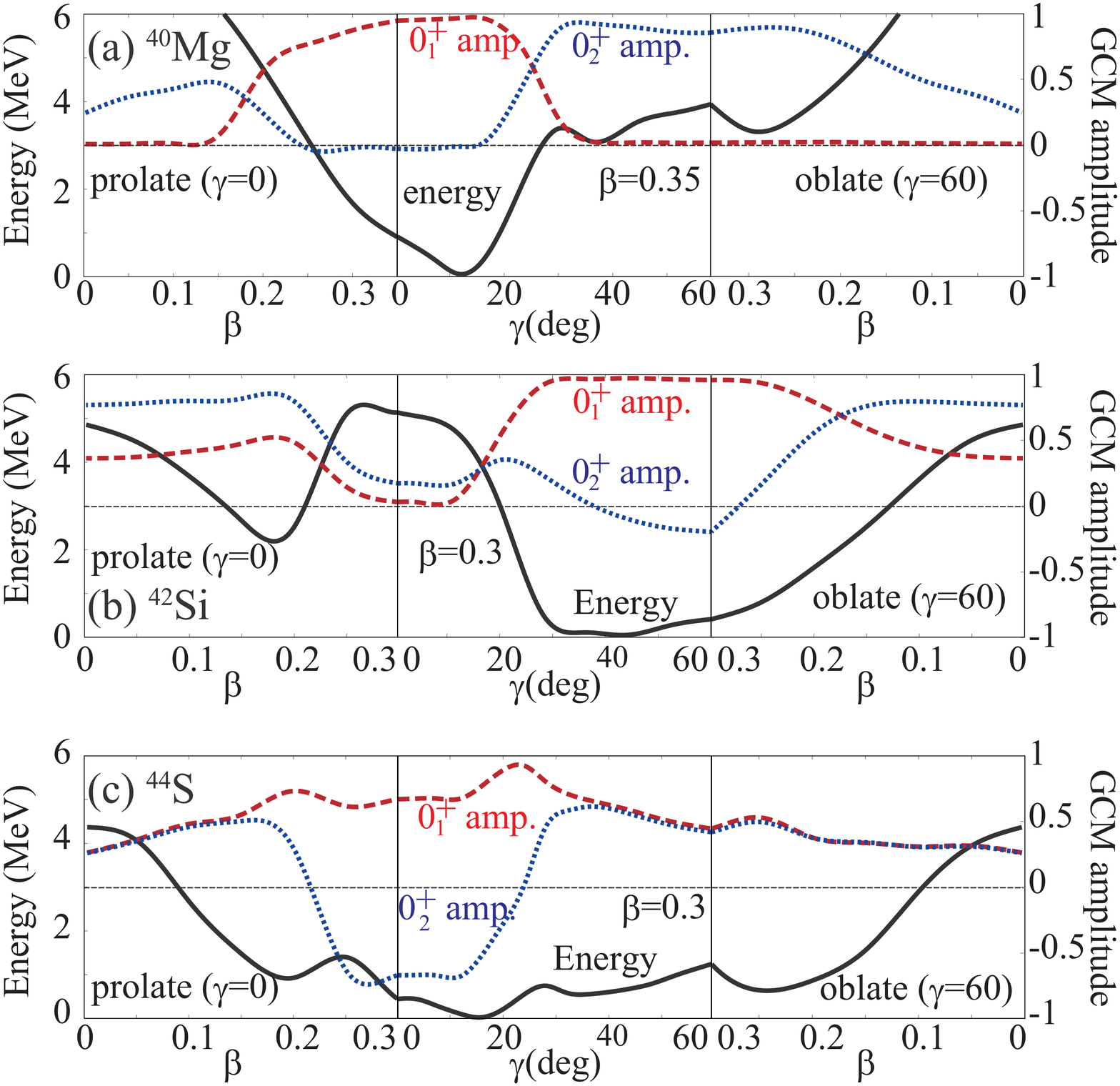}
\caption{Energy surface and the GCM amplitudes of the ground and $0^+_2$ states of 
 (a) $^{40}{\rm Mg}$, (b) $^{42}{\rm Si}$ and (c) $^{44}{\rm S}$, which are plotted as functions of
 $\beta$ and $\gamma$ along the sector path shown in Fig.~\ref{fig:surface}. The left and right
 panels correspond to the prolate ($\gamma=0^\circ$) and oblate ($\gamma=60^\circ$) shapes, whereas
 the middle panels show the axial asymmetric shape ($0^\circ\leq \gamma \leq 60^\circ$) with
 $\beta=0.35$ for $^{40}{\rm Mg}$ and  $\beta=0.3$ for $^{42}{\rm Si}$ and $^{44}{\rm S}$.} 
\label{fig:amp}
\end{figure}

\begin{table}[hbt]
  \caption{Reduced electric quadrupole transition probabilities in units of $e^2\rm fm^4$ and
 electric quadrupole moments in units of $\rm fm^2$. Experimental data is taken from
 Refs.~\cite{Glasmacher1997,Force2010,Longfellow2021}.} 
 \label{tab:e2} 
 \begin{ruledtabular}
  \begin{tabular}{cccccc}
   $B(E2; J_i\rightarrow J_f)$ & $^{40}{\rm Mg}$& $^{42}{\rm Si}$
	   & $^{44}{\rm S}$& Expt.($^{44}{\rm S}$)\\
   \hline
   $2^+_1\rightarrow 0^+_1$ & 97 & 73 & 77 & 63(18)\cite{Glasmacher1997}, 44(6)\cite{Longfellow2021}\\
   $2^+_2\rightarrow 0^+_2$ & 63 & --  & 71 & --\\
   $2^+_1\rightarrow 0^+_2$ & 0  & 7  & 4  & 8.4(26)\cite{Force2010}\\
   $2^+_2\rightarrow 0^+_1$ & 0  & --  & 2  & --\\
   $2^+_2\rightarrow 2^+_1$ & 2  & --  & 2  & --\\
  \end{tabular}
  \vspace{0.1cm}
  \begin{tabular}{cccc}
   $Q(2^+_n)$ & $^{40}{\rm Mg}$& $^{42}{\rm Si}$
	   & $^{44}{\rm S}$\\
   \hline
   $2^+_1$ & $-$20 & 17 & $-$18 \\
   $2^+_2$ & $-$6  & --  & $-$18 \\
  \end{tabular}
 \end{ruledtabular}
\end{table}

We see that $^{40}{\rm Mg}$ has the energy minimum at $(\beta, \gamma) = (0.36,14^\circ)$ in the
prolately-deformed region ($0<\gamma<30^\circ$). Correspondingly, the  squared GCM amplitude of the
ground state is  large around this energy minimum. In addition, $^{40}{\rm Mg}$ has a local energy
minimum at $(\beta,\gamma)=(0.31,44^\circ)$ in the oblately-deformed region 
($30^\circ <\gamma<60^\circ$), which is 2.8 MeV higher than the prolately-deformed minimum. The
squared GCM amplitude of the $0^+_2$ state is large around this local minimum. To illustrate the
situation more clearly, Fig.~\ref{fig:amp} (a) shows the energy and the GCM amplitude (not squared)
as a function of $\beta$ and $\gamma$ along the sector path shown in Fig.~\ref{fig:surface}. 
The GCM amplitude is large and localized in the prolately-deformed region  for the ground state,
whereas the $0_2^+$ state is localized in the oblately-deformed region. This indicates that both the
states have rigid shapes with small fluctuation. In other words,  the prolately- and
oblately-deformed rigid rotors coexist in the low-lying states of $^{40}{\rm Mg}$.  

This feature is well reflected in the reduced electric quadrupole transition probability ($B(E2)$), 
listed in Tab.~\ref{tab:e2}. Due to large prolate and oblate deformation, the in-band transitions
($2^+_1\rightarrow 0^+_1$ and $2^+_2\rightarrow 0^+_2$) are strong. On the contrary, the inter-band
transitions are one order of magnitude weaker than the in-band transitions,
because considerably different shapes of these states result in a small quadrupole matrix
element. The electric-quadrupole ($Q$) moment of the $2^+_1$ state exhibits a large negative value,
which is consistent with an estimation from a prolately-deformed rigid rotor
model~\cite{Bohr1969}. The $Q$ moment of the $2^+_2$ state  also has a negative value despite that
the state favors the oblately-deformed shape. It implies non-negligible contribution from the
fluctuation against $\gamma$ deformation. 

As seen in Fig.~\ref{fig:surface}, the nuclear shape of $^{42}{\rm Si}$ is different from
$^{40}{\rm Mg}$. The ground state is localized around the oblately-deformed energy minimum
at $(\beta,\gamma)=(0.31,44^\circ)$. Although there is no local minimum at the spherical shape,
the energy of the spherical state is relatively lower than deformed states. Consequently, the
squared GCM amplitudes of the $0^+_2$ state are distributed
in the region with $\beta < 0.2$,  and thus the $0^+_2$ state does not constitute a rotational
band. We also note that the $Q$ moment of the $2^+_1$ state is a large positive value, which is
consistent with an estimation from the oblately-deformed rigid rotor model~\cite{Bohr1969}.

Finally, we discuss $^{44}{\rm S}$, which shows an interesting aspect considerably different from   
$^{40}{\rm Mg}$ and $^{42}{\rm Si}$. The energy minimum shown in Fig.~\ref{fig:surface}
is located at $(\beta,\gamma)=(0.31, 16^\circ)$ but the energy is almost constant  against $\gamma$
deformation. As shown in the middle panel of Fig.~\ref{fig:amp}~(c),  when $\beta$ is fixed to 0.3,
the energy changes by only 1 MeV as a function of $\gamma$. Because of this flat energy surface,
the GCM amplitude of the ground state has a broad and non-localized distribution.
The same holds for the $0^+_2$ state, while it exhibits a node near $\gamma=30^\circ$ due to the
orthogonality to the ground state. These states may be regarded as ``large amplitude collective motion''.
In other words, $^{44}{\rm S}$ does not have any specific shape
but is always fluctuating.  We also note that the calculated
in-band and  intra-band $B(E2)$ values are consistent with the experimental data, although the
data have a large uncertainty.

\begin{figure}[ht]
\centering
\includegraphics[width=0.5\hsize]{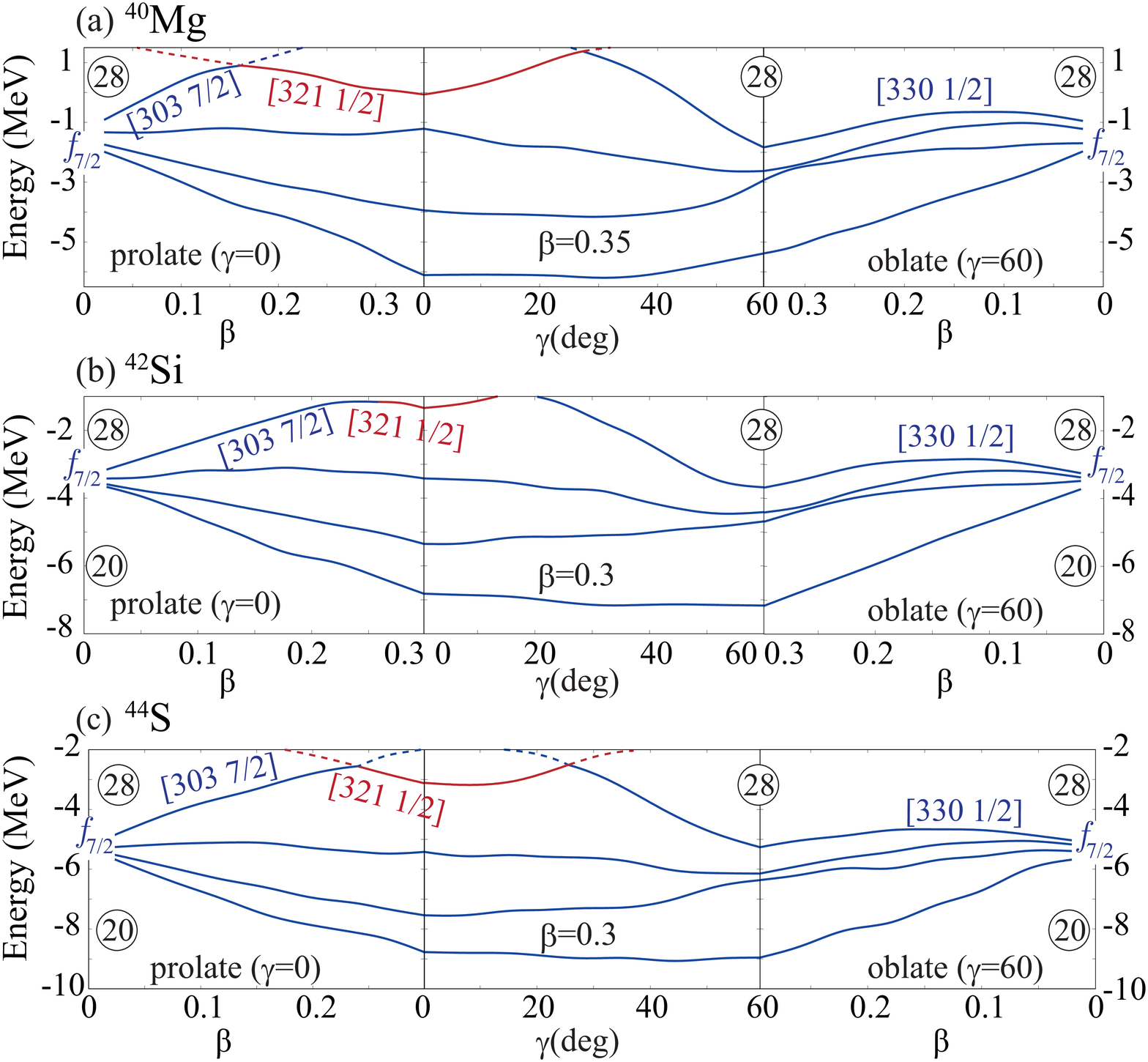}
\caption{Neutron single-particle energies of (a) $^{40}{\rm  Mg}$, (b) $^{42}{\rm Si}$ and (c)
  $^{44}{\rm S}$ as a function of $\beta$ and  $\gamma$ along the sector path shown in
 Fig.~\ref{fig:surface}.}\label{fig:spl}
\end{figure}

Thus,  various deformed states coexist in $^{40}{\rm Mg}$, $^{42}{\rm Si}$, and  $^{44}{\rm S}$. It
is important to note that the $N=28$ shell closure is lost in the ground states of all these nuclei,
and  the mechanism behind it depends on the nuclear shape of individual
nuclei~\cite{Suzuki2021}. Here, we  reiterate it for the discussion of the monopole transitions.
Figure~\ref{fig:spl} draws the neutron 
single-particle energies near the Fermi surface. In the prolately-deformed region, the neutron
orbits originating in the spherical $0f_{7/2}$ and $1p_{3/2}$ orbits are
inverted~\cite{Hamamoto2016}. More specifically, the ordering of the
$[Nn_z\Lambda\Omega]=[303\,7/2]$ and $[321\,1/2]$ orbits changes at around $\beta=0.2$--0.3, where  
$[Nn_z\Lambda\Omega]$ denotes the asymptotic quantum numbers of the Nilsson
orbit~\cite{Nilsson1955}. Thus, the $N=28$ shell gap disappears in the prolately-deformed ground  
state of  $^{40}{\rm Mg}$ due to the inversion of the neutron orbits.  

In the oblately-deformed states, the $N=28$ shell gap (the energy gap between the orbits originating
from the spherical $0f_{7/2}$ and $1p_{3/2}$ orbits) is kept large (see Fig.~\ref{fig:spl} (b).).
However, the orbits, which originate from the spherical $0f_{7/2}$ orbit and have the 
asymptotic quantum number $\Omega=1/2$ or 3/2 ([330\,1/2] and [321\,3/2]), are no longer the
eigenstate of the single-particle angular momentum but a mixed state with the $f$- and $p$-waves.
The $p$-wave mixing becomes stronger as the oblate deformation gets
larger~\cite{Hamamoto2016,Suzuki2021}.  Consequently, a couple of neutrons occupy the $p$-wave state
in the oblately-deformed ground state of $^{42}{\rm Si}$ (and in the $0^+_2$ state of $^{40}{\rm Mg}$)
despite that the energies of the neutron orbits shows a robust shell gap. We also note that the $f$-
and $p$-wave mixing does not occur in the prolately-deformed side as the [303\,7/2] orbit at the Fermi
level has $\Omega=7/2$.

In summary, the $N=28$ isotones show different aspects of the shape coexistence. In $^{40}{\rm Mg}$,
prolately- and oblately-deformed rigid rotors coexist, whereas an oblately-deformed rigid rotor and a
spherical state coexist in $^{42}{\rm Si}$. In contrast to these nuclei, $^{44}{\rm S}$  has no
specific shape, exhibiting large shape fluctuation. The difference in nuclear shape is strongly
correlated with the single-particle structure near the neutron Fermi surface.  In the
prolately-deformed states, the $N=28$ shell gap disappears due to the inversion of the neutron
orbits. On the other hand, in the oblately-deformed states, the shell gap is robust, but the $f$-
and $p$-wave mixing in the neutron orbits erodes the shell closure. Unexpectedly, the neutron
occupation numbers of the $p$-wave orbits, which is a measure for the robustness of the $N=28$ shell
closure, is not sensitive to the shape and neutron orbits of individual nuclei.  As calculated in
Ref.~\cite{Suzuki2021}, they are 2.0, 2,1 and 1.7 for $^{40}{\rm Mg}$, $^{42}{\rm Si}$ and
$^{44}{\rm S}$, respectively. Therefore, the neutron occupation numbers do not serve as a probe for
the nuclear shape. In the next subsection, we discuss the monopole transition strength can be a
probe for the nuclear shape rather than the occupation numbers.  
  
\subsection{Monopole transitions}

\begin{table}[hbt]
  \caption{Electric ($E0$) and isoscalar ($IS0$) monopole transition strengths between the
 $0^+_1$ and $0^+_2$ states in Weisskopf unit (Wu).} \label{tab:e0}  
 \begin{ruledtabular}
  \begin{tabular}{cccccc}
   & $^{40}{\rm Mg}$& $^{42}{\rm Si}$
	   & $^{44}{\rm S}$& Expt.($^{44}{\rm S}$)~\cite{Grevy2005}\\
   \hline
   $B(E0;0^+_2\rightarrow 0^+_1)$  & 0.0 & 0.2 & 0.04 & 0.022(2)\\
   $B(IS0;0^+_2\rightarrow 0^+_1)$ & 0.0 & 2.3 & 0.38 & --\\
  \end{tabular}
 \end{ruledtabular}
\end{table}
In this subsection, we consider the monopole transition strengths between the ground and $0^+_2$ states.
Electric ($E0$) and isoscalar ($IS0$) monopole transition operators are respectively defined by
\begin{align}
  \mathcal{M}_{E0}&=\sum_{i=1}^A r_i^2\frac{1+\tau_z}{2},\\
 \mathcal{M}_{IS0}&=\sum_{i=1}^A r_i^2,
\end{align}
where $r_i$ is the single-particle coordinate measured from the center-of-mass of the system. The
calculated reduced transition strengths from the ground state ($0_1^+$) to the $0_2^+$ state are
listed in Tab.~\ref{tab:e0}.  The transition strengths strongly depend on the nuclear shape:
In $^{40}{\rm Mg}$, the transition strengths are hindered and almost forbidden, whereas in
$^{42}{\rm Si}$ they are enhanced in order of Weisskopf unit. The strengths in $^{44}{\rm S}$ are
in between them and the calculated $B(E0)$ is close to the observed one~\cite{Grevy2005}.

To understand the origin of the different strengths of the monopole transitions,  let us review the
relationship between the shape  coexistence and monopole transitions~\cite{Heyde2011}. Suppose that
there are two state vectors $\ket{A}$ and $\ket{B}$ with different nuclear shapes, and the $0^+_1$
and $0^+_2$ states are described by their linear combinations 
\begin{align}
 \ket{0^+_1} &= a\ket{A} + b\ket{B},\\
 \ket{0^+_2} &= -b\ket{A} + a\ket{B},
\end{align}
where the coefficients $a$ and $b$ describe the mixing of two shapes. The monopole transition matrix
element can be written as 
\begin{align}
 \braket{0^+_2|\mathcal{M}|0^+_1} =& ab\left\{
 \braket{B|\mathcal M|B} - \braket{A|\mathcal M|A}\right\} \nonumber\\
 &+ (a^2-b^2)\braket{B|\mathcal M|A}, \label{eq:matrix}
\end{align}
where $\mathcal M$ is either $\mathcal{M}_{E0}$ or $\mathcal{M}_{IS0}$. The first term becomes large
when the radii of $\ket{A}$ and $\ket{B}$ are different, and the mixing is strong ($a\approx b
\approx 1/\sqrt{2}$). The second term vanishes when the particle(p)-hole(h) configurations of
$\ket{A}$ and $\ket{B}$  differ by more than 2p2h as $\mathcal{M}$ is a one-body operator. 
Keeping this in mind, we discuss the relationship between the monopole strength and shape
coexistence of the $N=28$ isotones. 

In $^{40}{\rm Mg}$, prolately- and oblately-deformed shapes coexist and their mixing is
small. Hence, we may take $\ket{A} = \ket{\rm prolate}$, $\ket{B} = \ket{\rm oblate}$, $a = 1$, 
$b = 0$. Then, Eq.~(\ref{eq:matrix}) reads 
\begin{align}
 \braket{0^+_2|\mathcal M|0^+_1} =
 \braket{\rm oblate|\mathcal M|\rm prolate}.\label{eq:matrixMg}
\end{align}
This matrix element vanishes due to the following reasons. As already explained, the
prolately-deformed ground state has two valence neutrons in an intruder orbit [321\,1/2] originating
in the spherical $1p_{3/2}$  orbit. Abbreviating the neutron orbits originating from the spherical
$0f_{7/2}$ ($1p_{3/2}$) orbit with the asymptotic quantum number $\Omega$ by $\ket{\Omega}$
($\ket{\overline{\Omega}}$), a dominant configuration of the eight valence neutrons may be written
as
\begin{align}
 \ket{\rm prolate}= \ket{1/2}^2\ket{3/2}^2\ket{5/2}^2\ket{\overline{1/2}}^2.\label{eq:prolate}
\end{align}
The first three orbits originate from the spherical $0f_{7/2}$  orbit, while the last one denoted by 
$\ket{\overline{1/2}}$ (the [321\,1/2]  orbit) is from the $1p_{3/2}$ orbit. In the case of
oblately-deformed shape, all valence neutrons occupy the orbits that originate from the spherical
$0f_{7/2}$ orbit, which is given by 
\begin{align}
 \ket{\rm oblate}= \ket{1/2}^2\ket{3/2}^2\ket{5/2}^2\ket{7/2}^2.\label{eq:oblate}
\end{align}
Thus, two configurations differ by at least 2p2h, i.e., the last two neutrons occupy the orbit
$\ket{\overline{1/2}}$ in Eq.~(\ref{eq:prolate}), while they occupy $\ket{7/2}$ in
Eq.~(\ref{eq:oblate}). Consequently, Eq.~(\ref{eq:matrixMg}) vanishes. We also note that the
monopole operator does not change the asymptotic quantum number $\Omega$, which is another reason
why the monopole transition between Eqs.~(\ref{eq:prolate}) and (\ref{eq:oblate}) is forbidden.

For $^{42}{\rm Si}$, we set  $\ket{A} = \ket{\rm oblate}$, $\ket{B} = \ket{\rm spherical}$, $a = 1$,
$b = 0$, and the transition matrix is given as 
\begin{align}
 \braket{0^+_2|\mathcal M|0^+_1} =
  \braket{\rm spherical|\mathcal M|\rm oblate}. \label{eq:matrixSi}
\end{align}
The oblately-deformed state is represented by Eq.~(\ref{eq:oblate}), and the
spherical state is also written as
\begin{align}
 \ket{\rm spherical}= \ket{1/2}^2\ket{3/2}^2\ket{5/2}^2\ket{7/2}^2.\label{eq:spherical}
\end{align}
Note that Eqs.~(\ref{eq:oblate}) and (\ref{eq:spherical}) mean that oblately-deformed and spherical
states are nonorthogonal, though they are not identical. These oblate and spherical configurations
smoothly transform as function of $\beta$ without level inversion, e.g., see
Fig.~\ref{fig:spl}~(b). Therefore, the monopole matrix element has a finite value and increases when
the overlap of two configurations is large.   

Differently from $^{40}{\rm Mg}$ and $^{42}{\rm Si}$, the $^{44}{\rm S}$ nucleus exhibits
large-amplitude collective motion. Let us approximate it as a mixture of prolate and oblate shapes
with equal amplitudes, i.e., $\ket{A}=\ket{\rm prolate}$, $\ket{B}=\ket{\rm oblate}$,
$a=b=1/\sqrt{2}$. In this case, the second term in Eq.~(\ref{eq:matrix}) vanishes and
the transition matrix is reduced to the difference of radii between the prolate and oblate shapes 
\begin{align}
 &\braket{0^+_2|\mathcal M|0^+_1} = \nonumber \\
 &\frac{1}{2}\left\{\braket{\rm oblate|\mathcal M|\rm oblate}
 -\braket{\rm prolate|\mathcal M|\rm prolate} \right\}.
 \label{eq:matrixS}
\end{align}
Eq.~(\ref{eq:matrixS}) gives a reasonable estimate of the transition strength. Applying the single
AMD wave functions with the deformation of $(\beta,\gamma) = (0.31,16^\circ)$ and $(0.23, 49^\circ)$ 
as $\ket{\rm prolate}$ and $\ket{\rm oblate}$, respectively, Eq.~(\ref{eq:matrixS}) yields 
$B(E0) =0.05$ Wu and $B(IS0) = 0.4$ Wu, which are not far from the results of the GCM calculation
listed in Tab.~\ref{tab:e0}.  

As we can see from these results, there is an interesting relationship between the monopole
strengths, shape coexistence and the neutron orbits near the Fermi level: In $^{40}{\rm Mg}$, the
valence neutron configurations of the prolately-deformed ground state and the oblately-deformed
$0_2^+$ state are different at least 2p2h due to the inversion of neutron orbits, and thus the
monopole transition is forbidden. In $^{42}{\rm Si}$, the neutron orbits in the oblately-deformed
ground state and the spherical $0_2^+$ state belong to the same class of the Nilsson orbits without
level inversion. Therefore, the monopole transition is enhanced. $^{44}{\rm S}$ exhibits the mixing
of the prolately- and oblately-deformed shapes, i.e., large-amplitude collective motion. Though the
monopole transition between the two different shapes is forbidden, the transition between the ground
and $0^+_2$ state can be strong, which is roughly proportional to the difference of the radii
between two shapes.

\section{summary}\label{sec:summary}

We have studied the structure of the neutron-rich $N=28$ isotones, $^{40}{\rm Mg}$, $^{42}{\rm Si}$
and $^{44}{\rm S}$, based on a fully microscopic framework of AMD. Our calculations reasonably
reproduced the observed data for the ground band, and predict the shape coexistence phenomena induced
by the quenching of the $N=28$ shell gap.

From the analysis of the energy surfaces and GCM amplitudes, we find that the spectra of $N = 28$
isotones show different aspects of shape coexistence. In $^{40}{\rm Mg}$, prolately- and
oblately-deformed rigid rotors coexist, and the $N = 28$ shell gap is lost due to the inversion
of the neutron orbits. On the other hand, an oblately-deformed rigid rotor and a spherical state
coexist in $^{42}{\rm Si}$ where the energy gap in the neutron orbits is robust. However the shell
closure is lost in the ground state because of the mixing of $f$ and $p$-wave near the Fermi
level. Differently from these nuclei,  $^{44}{\rm S}$ has large shape fluctuation and does not have
any definite shape, which can be regarded as large amplitude collective motion. 

We also point out that the neutron occupation number in the $p$-orbit is not sensitive to the
nuclear shape, and proposed the monopole transition strength as an alternative probe for the shape
coexistence phenomena in this mass region. In $^{40}{\rm Mg}$, the monopole transition strength from
the ground to first excited $0^+$ states is strongly hindered due to the inversion of neutron orbit,
while in $^{42}{\rm Si}$, the transition strength is significantly enhanced because these two states
do not have the inversion of neutron orbits. In $^{44}{\rm S}$, the large amplitude collective
motion yields intermediate monopole transition strength in the present AMD result. Such experimental
measurements will give us a deeper understanding of the shape coexistence phenomena and the erosion
of the $N=28$ shell closure in this mass region.  

\begin{acknowledgements}
 The authors thank to Prof. K. Yoshida and K. Washiyama for the helpful discussions. 
 This work was in part supported by JSPS KAKENHI Grants Nos. 18K03635 and 19K03859. 
 Part of the numerical calculations were performed using Oakforest-PACS at the Center for
 Computational Sciences in the University of Tsukuba. We acknowledge the collaborative research
 program 2021, Information Initiative Center, Hokkaido University.
\end{acknowledgements}

\bibliography{export}

\end{document}